\documentclass[12pt]{article}

\usepackage{epsf}
\usepackage[dvips]{graphicx}

\begin{document}

\begin{center}
{\bfseries MODEL INDEPENDENT CONSIDERATIONS ON TWO-PHOTON EXCHANGE IN ELECTRON PROTON SCATTERING AND IN THE CROSSED PROCESSES
}

\vskip 5mm

E. Tomasi-Gustafsson
\vskip 5mm

{\small
{\it
DAPNIA/SPhN- CEA/SACLAY, 91191 Gif-sur-Yvette cedex France}
}
\\
{\it
E-mail: etomasi@cea.fr
}
\end{center}

\vskip 5mm

\begin{center}
\begin{minipage}{150mm}
\centerline{\bf Abstract}

The presence of two photon exchange in $ep$ elastic scattering and in the crossed processes ($\bar p +p\leftrightarrow \ell^+ +\ell^-$, $\ell=e$ or $\mu$) is discussed  in terms of three complex amplitudes which are functions of two kinematical variables, the momentum transfer squared $q^2$ and the polarization of the virtual photon, $\epsilon$. Model independent expressions of the differential cross section and of polarization observables are derived. Particular attention is devoted to the extraction of hadron electromagnetic form factors which becomes much more difficult in presence of two photon exchange, and to the relevant experimental observables which could give evidence for such mechanism. 
\end{minipage}
\end{center}

\section{Introduction}

Electron hadron scattering is considered to be the best way to study the hadron structure. The underlying mechanism in elastic (and inelastic) scattering is one photon exchange which can be exactly calculated in quantum electrodynamics (QED), at least for the lepton vertex.  In this respect the electromagnetic probes are traditionally preferred to the hadronic beams for the investigation of nucleons and light nuclei \cite{Ho62}. Radiative corrections, firstly calculated by Schwinger \cite{Shwinger}, should be taken precisely into account, and are very important, in particular for the experimental determination of the differential cross section of $ep$ scattering. 

Due to the steep decreasing of the electromagnetic form factors (FFs)  when  the momentum transfer squared, $Q^2$, increases, another mechanism, the two photon (TPE) exchange mechanism, where $Q^2$ is equally shared between the two photons, can become important. This fact was already indicated in the seventies \cite{Gu73}, but it was never experimentally observed. Numerous tests of the validity of the one photon mechanism have been done in the past, using different methods: test of the linearity of the Rosenbluth formula for the differential cross section, comparison of the $e^+p$ and $e^-p$-cross sections, attempts to measure various T-odd polarization observables, but no effect was visible beyond the precision of the experimental data. Only recently, measurements of the asymmetry in the scattering of transversely polarized
electrons on unpolarized protons found values different from zero, contrary
to what is expected in the Born approximation \cite{We01},

Note that the TPE exchange should appear at smaller $Q^2$ for heavier targets: $d$, $^3\!He$,  $^4\!He$, because the corresponding FFs decrease faster with $Q^2$ in comparison with protons. In \cite{Re99} the possible effects of TPE  have been estimated from the precise data on the structure function $A(Q^2)$, obtained at JLab in electron deuteron elastic scattering, up to $Q^2=6$ GeV$^2$ \cite{Al99}. The possibility of a  $2\gamma$ contribution  has not been excluded by this analysis, starting from $Q^2=1$ GeV$^2$, and the necessity of  dedicated experiments was pointed out. From this kind of considerations, one would expect to observe the  TPE contribution in $eN$-scattering at larger momentum transfer, for $Q^2\simeq 10$ GeV$^2$. In Refs. \cite{Re04}, on the line of the $ed$-analysis \cite{Re99}, we proved that general properties of the hadron electromagnetic interaction, such as the C-invariance and the crossing symmetry, give rigorous prescriptions for different observables for the elastic scattering of electrons and positrons by nucleons. Model independent prescriptions are very helpful in identifying possible manifestations of the TPE exchange mechanism. 

Recent developments in the field of hadron electromagnetic FFs are due to the very precise and surprising data obtained at the Jefferson Laboratory (JLab), in $\vec e+p\to e+\vec p$ elastic scattering \cite{Jo00}, based on the polarization transfer method \cite{Re68}, which show that the electric and magnetic distributions in the proton are different, contrary to what was previously assumed.

The present data show a discrepancy between the $Q^2$-dependence of the ratio
$R= \mu_p G_{Ep}$/$G_{Mp}$ of the electric to the magnetic proton FFs ($\mu_p$=2.79 is the proton magnetic moment), whether derived with the standard Rosenbluth separation \cite{Ar75} or with the polarization method. 

An important point is the calculation of radiative corrections. These corrections are large (in absolute value) for the differential cross section \cite{Mo69}, in particular for high resolution experiments, whereas they are negligible for the ratio $P_L/P_T$ of longitudinal to transverse polarization of the proton emitted in the elastic collision of polarized electrons with an unpolarized proton target.

A careful experimental and theoretical analysis of this problem is necessary, including the investigation of the  TPE contribution. We present a model independent strategy to extract electromagnetic nucleon FFs and to determine the  TPE contribution, on the basis of a general analysis of polarization phenomena in elastic $eN$-scattering  \cite{Re04} and in the crossed channels \cite{Ga05,Ga06}. We discuss also the possibility to solve this problem, having a positron beam. We show that the measurement of the differential cross section and of the components of the proton polarization in $e^-p$ and $e^+p$ elastic scattering (in identical kinematical conditions) is the most direct way to access the nucleon FFs and the TPE contribution. In absence of positron beam, an alternative way is the measurements of a larger set of polarization observables in $e^-p$-scattering.  

In the time--like (TL) region of momentum transfer, investigated with annihilation processes, FFs are complex functions of $Q^2$. They can be determined through an angular distribution measurement, at fixed energy. This is, in principle, easier than the Rosenbluth separation, where each measurement corresponds to a fixed $Q^2$, and requires changing the initial energy and the scattering angle. Due to the available beam intensity, the statistics has not been sufficient up to now, for the individual extraction of both FFs, and the data are presented under the assumption $|G_E|=|G_M|$ or $|G_E|=0$.

The reaction $\bar p + p\to \ell^++\ell^-$,
$\ell=e$ or $\mu $  was firstly considered in Ref. \cite{Zi62} in the case
of unpolarized particles. As in the space--like (SL) region, the TPE can also become important in TL region if the nucleon FFs decrease rapidly with $q^2$. Moreover, as the $2\gamma$ amplitude is expected to be mostly imaginary, it will contribute in TL region, where FFs are complex. The general case of polarized initial particles (antiproton beam or/and proton target) in $\bar p + p\to e^++e^-$ has been firstly investigated in Ref. \cite{Bi93}, with particular attention to the determination of the
phases of FFs, and more recently in Ref. \cite{ETG05}. 

In this talk we present general expressions for polarization observables and indicate model independent methods to extract nucleon electromagnetic FFs in presence of TPE. The parametrization of the $2\gamma$ term can be done in tensor or axial forms. It is possible to show that the expressions for the observables are exactly the same, whether for generalized FFs, for example, they depend on the parametrization used.

\section{Scattering channel}
The starting point of our analysis is the following general parametrization  of the spin structure of the matrix element for elastic $e^{\pm}N$-scattering, according to the formalism of \cite{Go57}:
\begin{equation}
{\cal  M}=\displaystyle\frac{e^2}{Q^2}\overline{u}(k_2)\gamma_{\mu}u(k_1)
\overline{u}(p_2) \left [{\cal  A}_1(s,Q^2)\gamma_{\mu}-{\cal  A}_2(s,Q^2) 
\displaystyle\frac{\sigma_{\mu\nu}q_{\nu}}{2m}+{\cal  A}_3(s,Q^2)\hat K {\cal  P}_{\mu}\right] u(p_1),
\label{eq:mat1}
\end{equation}
where $K=(k_1+k_2)/2$, ${\cal  P}=(p_1+p_2)/2$, ${\cal  A}_1-{\cal  A}_3$ are the corresponding invariant amplitudes, $k_1$ $(p_1)$ and $k_2$ $(p_2)$ are the four-momenta of the initial and final electron (nucleon), $m$ is the nucleon mass, $q=k_1-k_2$, $Q^2=-q^2>0$.
In case of one-photon exchange 
$$
{\cal  A}_1(s,Q^2)\to F_{1}(Q^2),~
{\cal  A}_2(s,Q^2)\to F_{2}(Q^2),~
{\cal  A}_3\to 0.
$$
$F_{1}$ and $F_{2}$ are the Dirac and Pauli nucleon electromagnetic FFs, which are real functions of the variable $Q^2$ - in the SL region of momentum transfer. 

Taking into account the identity of the initial and final states and the T-invariance of the electromagnetic interaction, the processes $e^{\mp}N\to e^{\mp} N$, in which  four  particles with spin 1/2 participate, are characterized by six independent products of four-spinors, describing the initial and final fermions.  The corresponding (model independent) parametrization of the matrix element can be done in many different but equivalent forms, in terms of six invariant complex amplitudes, ${\cal A}_i(s,Q^2)$, $i=1-6$, which are functions of two independent variables,  and  $s=(k_1+p_1)^2$ is the square of the total energy of the colliding particles. 

In principle, another set of variables can be considered: $\epsilon$ and $Q^2$, which is equivalent to $s$ and $Q^2$: $\epsilon^{-1}=1+2(1+\tau)\tan ^2(\theta_e/2),$ where $\theta_e$ is the electron scattering angle in the laboratory (Lab) system, $\tau=Q^2/(4m^2)$. The variables $\epsilon$ and $Q^2$ are well adapted to the description of the properties of one photon exchange for elastic $eN$-scattering, because, in this case, only the $Q^2$-dependence of FFs has a dynamical origin, whereas the linear $\epsilon$-dependence of the differential cross section is a trivial consequence of the one photon mechanism. On the other hand, the variables $s$ and $Q^2$ are
better suited to the analysis of the implications from crossing symmetry.

The conservation of the lepton helicity, which is a general property of QED at high energies, reduces the number of invariant amplitudes for elastic $eN$-scattering,  from six to three.

In the general case (with multiphoton exchanges) the situation is more complicated, because the amplitudes ${\cal  A}_i(s,Q^2)$, $i=1-3$, are complex functions of two independent variables, $s$ and $Q^2$; moreover, their  connection with the nucleon electromagnetic FFs, $F_{i}(Q^2)$, is non-trivial, because these amplitudes depend on many quantities, as, for example, the FFs of the $\Delta$-excitation - through the amplitudes of the virtual Compton scattering. Electron and positron scattering are no more described by the same set of amplitudes.

In this framework, the simple and transparent phenomenology of electron-hadron physics does not hold anymore, and in particular, it would  be very difficult  to extract information on the internal structure of a hadron in terms of electromagnetic FFs, which are real functions of one variable, from electron scattering experiments.

We use the following notations (the sign $-(+)$ indicates electron(positron) sctattering:
\begin{eqnarray*}
\widetilde{G}_E^{(\mp)}(Q^2,\epsilon)&=&{\cal  A}_1^{(\mp)}-\tau {\cal  A}_2^{(\mp)}(Q^2,\epsilon),\\
\widetilde{G}_M^{(\mp)}(Q^2,\epsilon)&=&{\cal  A}_1^{(\mp)}+{\cal  A}_2^{(\mp)}(Q^2,\epsilon), 
\end{eqnarray*}
considering $\widetilde{G}_{E,M}^{(\mp)}(Q^2,\epsilon)$ as  {\it generalized} FFs, so that
\begin{equation}
\widetilde{G}_{E,M}^{(\mp)}(Q^2,\epsilon)=\mp G_{E,M}(Q^2)+\Delta G_{E,M}(Q^2,\epsilon).
\label{eq:gem}
\end{equation}
with $G_E(Q^2)=F_1(Q^2)-\tau F_2(Q^2)$, $G_M(Q^2)=F_1(Q^2)+F_2(Q^2)$.

Both reactions $e^{\mp}+N\to e^{\mp}+N$ are fully described by eight different real quantities: two real FFs $G_{E,M}(Q^2)$, which are functions of one variable only, and three  functions: $\Delta G_E(Q^2,\epsilon)$,$\Delta G_M(Q^2,\epsilon)$,${\cal A}_3(Q^2,\epsilon)$, which are, generally, complex functions of two variables, $Q^2$ and $\epsilon$. Therefore, these eight real functions  completely determine  six complex amplitudes 
${\cal A}_i^{(\mp)}(Q^2,\epsilon)$ 
for $e^{\mp}+N\to e^{\mp}+N$, therefore there are very special connections of 12 real functions $Re{\cal A}_i^{(\mp)}(Q^2,\epsilon)$ and 
$Im{\cal A}_i^{(\mp)}(Q^2,\epsilon)$, with 8 real functions:
\begin{equation}
{\cal A}_i^{(+)}(Q^2,\epsilon)-{\cal A}_i^{(-)}(Q^2,\epsilon)=2 F_i(Q^2),~i=1,2
\label{eq:am1}
\end{equation}
so:
\begin{equation}
Im[{\cal A}_i^{(+)}(Q^2,\epsilon)-{\cal A}_i^{(-)}(Q^2,\epsilon)]=0,~i=1,2
\label{eq:am2}
\end{equation}
in the whole region of SL momentum transfer.

Note that the C-invariance and the crossing symmetry require that the $\epsilon$-dependence of the six above quoted functions occurs through the argument: $x=\sqrt{(1+\epsilon)/(1-\epsilon)}$, with the following symmetry properties, with respect to the change $x\to -x$:
\begin{equation}
\Delta G_{E,M}(Q^2,-x)= -\Delta G_{E,M}(Q^2,x),~
{\cal A}_3(Q^2,-x)={\cal A}_3(Q^2,x).
\end{equation}
These expressions contain the physics of the nucleon electromagnetic structure, taking into account the $1\gamma\bigotimes2\gamma$-interference contribution. In this case, the complete experiment for $eN$-elastic scattering requires six additional functions, depending on two kinematical variables, instead of two real functions of a single variable $Q^2$. So, if previously the measurement of the differential cross section, with unpolarized particles in initial and final states, was in principle sufficient (through the Rosenbluth fit, linear in the variable $\epsilon$), now measurements with electrons and positrons in the same kinematical conditions are necessary. 

The FFs $G_{EN}(Q^2)$ and $G_{MN}(Q^2)$ and the $2\gamma$-amplitudes, $\Delta G_{E,M}(Q^2,\epsilon)$ are the same for $e^+p$ and $e^-p$ elastic scattering. This allows to connect the difference of the differential cross sections for $e^{\mp} p$-interaction with the deviations from the $\epsilon$-linearity of the Rosenbluth plot.

The $\epsilon$-dependence of the interference contribution to the differential cross section of $e^{\mp}p$ elastic scattering is very particular. Any approximation of this term by a linear function in the variable $\epsilon$ is in contradiction with C-invariance and crossing symmetry of the electromagnetic interaction. 

In absence of positron beam two other possibilities to measure $G_{E,M}(Q^2)$ can be suggested, using only an electron beam.

One possibility is the measurement of T-odd polarization observables, such as $P_y$, $D_{zy}(\lambda_e)$, and $D_{yz}(\lambda_e)$ (i.e., the components of the depolarization tensor, in the scattering of longitudinally polarized electrons by a polarized target, with the measurement of the final proton polarization). All these observables, which vanish in the Born approximation for $eN$-scattering, are of the order of $\alpha$ and should be measured with corresponding accuracy.

Another possible way requires the measurement of five T-even polarization observables (five quadratic combinations of three complex amplitudes), as  $d\sigma/d\Omega_e$, $P_x$ (or $A_x$), and the $D_{xx}$, $D_{yy}$ and $D_{xz}$ components of the depolarization tensor (for unpolarized electron scattering).

\section{Annihilation channel}

The analysis of annihilation channel follow the same steps as for elastic scattering. Therefore we focus here on some interesting polarization observables. Complete formulas and derivation can be found in  \cite{Ga05,Ga06}.
The differential cross section of the reaction $ e^++e^-\rightarrow N+\bar N$, 
for the case of unpolarized 
particles, has the form 
\begin{equation}\label{eq:eq13}
\frac{d\sigma_{un}}{d\Omega } = \frac{\alpha^2\beta }{4q^2}D, \
\end{equation}
with
$$D=(1+\cos^2\theta )(|G_{MN}|^2+2ReG_{MN}\Delta G_{MN}^*)+\frac{1}{\tau }
\sin^2\theta (|G_{EN}|^2+ $$
\begin{equation}
+2ReG_{EN}\Delta G_{EN}^*)- 
\frac{4}{\tau }\sqrt{\tau (\tau -1)}\cos\theta Re G_{MN}A_N^*, 
\label{eq:eq13a}
\end{equation}
where $\theta $ is the angle between the momenta of the electron and the detected antinucleon and $\beta $ is the nucleon velocity in CMS. Note that Eq. (\ref{eq:eq13}) was 
obtained neglecting the terms of the order of $\alpha ^2$ compared to the 
dominant (Born approximation) terms. In the one photon exchange limit the 
expression  (\ref{eq:eq13}) coincides with the result obtained for the differential cross section in Ref. \cite{Du96}. The TPE contribution brings three new terms of the order of $\alpha $ compared to the Born contribution. 

At the threshold of the reaction, $q^2 = 4m^2,$ the equality $G_{MN}=G_{EN}=G_N$ holds and Eq. (\ref{eq:eq13a}) reduces to 
$$D=D^{th}=|G_{N}|^2+ReG_{N}(\Delta G_{MN}^*+\Delta G_{EN}^*)+
\cos^2\theta ReG_{N}(\Delta G_{MN}^*-\Delta G_{EN}^*). $$

Symmetry properties of the amplitudes
with respect to the $\cos\theta \to -\cos\theta $ transformation can be derived from $ C $  invariance :
\begin{equation}
\Delta G_{MN,EN}(\cos\theta )=-\Delta G_{MN,EN}(-\cos\theta ),\ A_N(\cos\theta )=A_N(-\cos\theta ).
\label{eq:eq14}
\end{equation}
If the experiment does not distinguish the nucleon 
from the antinucleon, then the following sum of the differential cross sections is measured:
$$\frac{d\sigma_+}{d\Omega}=\frac{d\sigma}{d\Omega}(\cos\theta )+\frac{d\sigma}{d\Omega}(-\cos\theta ),  $$
which does not depend on the TPE 
terms. Moreover,  the total cross section is also independent of the TPE terms:
\begin{equation}\label{eq:eq15}
\sigma _t(q^2)=\frac {4\pi }{3}\frac {\alpha ^2\beta }{q^2}\left [|G_{M}(q^2)|^2+\frac{1}{2\tau }|G_{E}(q^2)|^2
\right ]. 
\end{equation} 
On the other hand, the relative contribution of TPE mechanism is enhanced in the following   angular asymmetry 
\begin{equation}\label{eq:eq16}
A(q^2,\theta_0)=\frac {\sigma (q^2,\theta_0)-\sigma (q^2,\pi -\theta_0)}
{\sigma (q^2,\theta_0)+\sigma (q^2,\pi -\theta_0)}, 
\end{equation}  
with $\sigma (q^2,\theta_0)=\int_0^{\theta_0}\frac{d\sigma }{d\Omega}(q^2,\theta )d\Omega , \ \ 
\sigma (q^2,\pi -\theta_0)=\int_{\pi -\theta_0}^{\pi }\frac{d\sigma }{d\Omega}(q^2,\theta )d\Omega . $
Using the symmetry relations (\ref{eq:eq14}) one obtains the  following expression
\begin{eqnarray}
&&A(q^2, \theta_0)=\frac{2}{d}\int_0^{\theta_0}dcos\theta \Bigg [(1+cos^2\theta )ReG_{MN}(q^2)\Delta G_{MN}^*(q^2,cos\theta )
\nonumber \\
&&\hspace{-1truecm}+\frac{sin^2\theta }{\tau }ReG_{EN}(q^2)\Delta G_{EN}^*(q^2,cos\theta )-\frac{2}{\tau }\sqrt{\tau (\tau -1)}cos\theta ReG_{MN}(q^2)A_N^*(q^2,cos\theta )\Bigg ],
\label{eq:eq17}
\end{eqnarray}
$\mbox{with} d=\frac{1-x_0}{3}\Bigg [(4+x_0+x_0^2)|G_{MN}|^2+\frac{1}{\tau }(2-x_0-x_0^2)|G_{EN}|^2 \Bigg ], \ \ x_0=cos\theta_0. $

The TPE contributions can be removed considering the sum of the quantities $\sigma (q^2,\theta_0)$ 
and $\sigma (q^2,\pi -\theta_0)$: 
\begin{equation}
\Sigma (q^2, \theta_0)=\sigma (q^2,\theta_0)+\sigma (q^2,\pi -\theta_0)=\frac{\pi \alpha^2}{q^2}\beta d.
\label{eq:eq18}
\end{equation}
always neglecting the terms of the order of $\alpha ^2$ with respect to the leading ones.

Note that, unlike elastic electron--nucleon scattering in the Born approximation, the hadronic 
tensor in the TL region contains a symmetric part 
even in the Born approximation due to the complexity of the nucleon FFs. 
Taking into account the TPE contribution leads to antisymmetric terms in 
this tensor, which induce non--zero polarization 
of the outgoing antinucleon $P_y$ (the initial state is unpolarized): 
\begin{eqnarray}
P_y&=&\frac{2sin\theta }{\sqrt{\tau }D}
\Bigg \{ cos\theta \left [
ImG_{MN}G_{EN}^*+Im(G_{MN}\Delta G_{EN}^*-G_{EN}\Delta G_{MN}^*)\right ]-\nonumber \\
&&-\sqrt{\frac{(\tau -1)}{\tau }}ImG_{EN}A_N^*\Bigg  \}.
\label{eq:eq20}
\end{eqnarray}
In the one--photon--exchange (Born) approximation this expression coincides with the result of  Ref. \cite{Du96}. $P_y$ is determined
by the polarization component which is perpendicular to the reaction plane and, being T--odd quantity, does not vanish even in the
one--photon--exchange approximation due to the complexity of the nucleon FFs
in the TL region. This is principal difference with the elastic
electron--nucleon scattering. In the Born approximation this polarization becomes equal to zero at the
scattering angle $\theta = 90^0$ (as well at $\theta = 0^0$ and $180^0$).
The presence of TPE leads to a non--zero value of the
polarization at this angle, which is expected to be of the order of $\alpha $. The measurement of this polarization at
$\theta = 90^0$ contains information about TPE and its
behavior as a function of $q^2.$

In the threshold region  this polarization vanishes,  in the Born approximation
due to the relation $G_{EN}=G_{MN}$ . TPE induces a non zero polarization. The effect of TPE for the polarization at an arbitrary scattering angle is expected to increase as $q^2$ increases, as the $2\gamma$ amplitudes decrease more slowly with $q^2$ in comparison with the nucleon FFs.

\section{Conclusions}

The general symmetry properties of electromagnetic interaction, such as the C-invariance, the crossing symmetry and the lepton helicity conservation in QED, allow to obtain  rigorous results concerning two-photon exchange contributions for elastic $eN$-scattering and to analyze the effects of this mechanism in $eN$-phenomenology \cite{Re04}.
We analyzed above different possible strategies in the determination of the nucleon electromagnetic FFs, $G_E(Q^2)$ and $G_M(Q^2)$, through the measurement of different polarization observables in elastic $e^{\pm} N$ scattering, of T-even and T-odd nature, in presence of TPE.

There are in principle three different ways , to determine the physical nucleon FFs, $G_{E,M}(Q^2)$. The proposed methods are, all, relatively complicated but allow to go beyond the description in terms of 'generalized' FFs, which are functions of two kinematical variables, $Q^2$ and $\epsilon$ and are not directly related to the nucleon electromagnetic structure.

The formally simplest way needs the parallel study of positron and electron scattering, in the same kinematical conditions. We showed that the two-photon contribution cancels in the sum of the differential cross sections, $d\sigma^{(-)}/d\Omega_e+d\sigma^{(+)}/d\Omega_e$. A linear $\epsilon$-fit of this quantity allows to extract $G_E(Q^2)$ and $G_M(Q^2)$, through a generalized Rosenbluth separation. At higher $Q^2$, due to the small contribution of $G_E(Q^2)$,  the polarization transfer method should be used, which requires the measurement of the $P_x$ and $P_z$-components of the final nucleon polarization -with longitudinally polarized electron and positron beams. This can be in principle realized at the HERA $e^{\pm}$ ring, with a polarized jet proton target.

In absence of a positron beam one has to  measure 3  T-odd polarization observables, such as $P_y$, $D_{zy}(\lambda_e)$, and $D_{yz}(\lambda_e)$ which are of the order of $\alpha$ or five T-even polarization observables, as $d\sigma/d\Omega_e$, $P_x$ (or $A_x$), and $D_{xx}$, $D_{yy}$ and $D_{xz}$.

Complete expressions and properties can be found in Refs. \cite{Re04} for the scattering channel and in Refs. \cite{Ga05,Ga06} for the annihilation channels.

Therefore, in presence of TPE, the extraction of the nucleon electromagnetic FFs is still possible, but requires more complicated experiments, with a very high level of precision. Only in this way it will be possible to investigate the nucleon structure, at large momentum transfer, keeping the elegant formalism of QED, traditionally used for this aim.

This work was initiated in collaboration with Prof. M. P. Rekalo. Thanks are due to G.I. Gakh for careful reading of the manuscript and valuable discussions.

{}
\end{document}